\documentclass[aps,prl,twocolumn,superscriptaddress]{revtex4-1}
\usepackage{graphicx}
\usepackage[rightcaption]{sidecap}
\usepackage{epstopdf}

\DeclareGraphicsExtensions{.pdf,.png,.jpg}

\usepackage{color}

\begin{document}

\title{A quantum magnetic analogue to the critical point of water}

\author{J. Larrea Jim\'enez}
\affiliation{Institute of Physics, University of S\~{a}o Paulo, CEP 05508-090, 
S\~{a}o Paulo, SP, Brazil}
\affiliation{Institute of Physics, Ecole Polytechnique F\'ed\'erale de 
Lausanne (EPFL), CH-1015 Lausanne, Switzerland}
\author{S. P. G. Crone}
\affiliation{Institute for Theoretical Physics and Delta Institute for 
Theoretical Physics, University of Amsterdam, Science Park 904, 1098 XH 
Amsterdam, The Netherlands}
\author{E. Fogh}
\affiliation{Institute of Physics, Ecole Polytechnique F\'ed\'erale de 
Lausanne (EPFL), CH-1015 Lausanne, Switzerland}
\author{M. E. Zayed}
\affiliation{Department of Physics, Carnegie Mellon University in Qatar, 
Education City, PO Box 24866, Doha, Qatar}
\author{R. Lortz}
\affiliation{Department of Physics, Hong Kong University of Science and 
Technology, Clearwater Bay, Kowloon, Hong Kong}
\author{E. Pomjakushina}
\affiliation{Laboratory for Multiscale Materials Experiments, Paul Scherrer 
Institute, CH-5232 Villigen-PSI, Switzerland.}
\author{K. Conder}
\affiliation{Laboratory for Multiscale Materials Experiments, Paul Scherrer 
Institute, CH-5232 Villigen-PSI, Switzerland.}
\author{A. M. L\"auchli}
\affiliation{Institut f\"ur Theoretische Physik, Universit\"at Innsbruck, 
A-6020 Innsbruck, Austria}
\author{L. Weber}
\affiliation{Institut f\"ur Theoretische Festk\"orperphysik, JARA-FIT and 
JARA-HPC, RWTH Aachen University, 52056 Aachen, Germany}
\author{S. Wessel}
\affiliation{Institut f\"ur Theoretische Festk\"orperphysik, JARA-FIT and 
JARA-HPC, RWTH Aachen University, 52056 Aachen, Germany}
\author{A. Honecker}
\affiliation{Laboratoire de Physique Th\'eorique et Mod\'elisation, CNRS 
UMR 8089, CY Cergy Paris Universit\'e, 95302 Cergy-Pontoise Cedex, France}
\author{B. Normand}
\affiliation{Paul Scherrer Institute, CH-5232 Villigen-PSI, Switzerland}
\affiliation{Institute of Physics, Ecole Polytechnique F\'ed\'erale de 
Lausanne (EPFL), CH-1015 Lausanne, Switzerland}
\author{Ch. R{\"u}egg}
\affiliation{Paul Scherrer Institute, CH-5232 Villigen-PSI, Switzerland}
\affiliation{Institute for Quantum Electronics, ETH Z\"urich, CH-8093 
H\"onggerberg, Switzerland}
\affiliation{Institute of Physics, Ecole Polytechnique F\'ed\'erale de 
Lausanne (EPFL), CH-1015 Lausanne, Switzerland}
\affiliation{Department of Quantum Matter Physics, University of Geneva, 
CH-1211 Geneva 4, Switzerland}
\author{P. Corboz}
\affiliation{Institute for Theoretical Physics and Delta Institute for 
Theoretical Physics, University of Amsterdam, Science Park 904, 1098 XH 
Amsterdam, The Netherlands}
\author{H. M. R\o{}nnow }
\affiliation{Institute of Physics, Ecole Polytechnique F\'ed\'erale de 
Lausanne (EPFL), CH-1015 Lausanne, Switzerland}
\author{F.~Mila}
\affiliation{Institute of Physics, Ecole Polytechnique F\'ed\'erale de 
Lausanne (EPFL), CH-1015 Lausanne, Switzerland}

\date{\today}

\maketitle

{\bf At the familiar liquid-gas phase transition in water, the density jumps 
discontinuously at atmospheric pressure, but the line of these first-order 
transitions defined by increasing pressures terminates at the critical 
point \cite{rcdlt}, a concept ubiquitous in statistical thermodynamics 
\cite{rcl}. In correlated quantum materials, a critical point was predicted 
\cite{Georges96} and measured \cite{Limelette03,Kanoda05} terminating 
the line of Mott metal-insulator transitions, which are also first-order
with a discontinuous charge density. In quantum spin systems, continuous 
quantum phase transitions (QPTs) \cite{rsbook} have been investigated 
extensively \cite{Rueegg08,Merchant14,Giamarchi08,Thielemann09,Yu12}, but 
discontinuous QPTs have received less attention. 
The frustrated quantum antiferromagnet SrCu$_2$(BO$_3$)$_2$ constitutes 
a near-exact realization of the paradigmatic Shastry-Sutherland model 
\cite{ShaSu81,MiUeda03,Corboz13} and displays exotic phenomena including 
magnetization plateaux \cite{Matsuda13}, anomalous thermodynamics 
\cite{Wietek19} and discontinuous QPTs \cite{Zayed17}.
We demonstrate by high-precision specific-heat measurements under pressure 
and applied magnetic field that, like water, the pressure-temperature phase 
diagram of SrCu$_2$(BO$_3$)$_2$ has an Ising critical point terminating a 
first-order transition line, which separates phases with different densities 
of magnetic particles (triplets).
We achieve a quantitative explanation of our data by detailed numerical 
calculations using newly-developed finite-temperature tensor-network methods 
\cite{Verstraete04,Jordan08,Czarnik19,Wietek19}. 
These results open a new dimension in understanding the thermodynamics 
of quantum magnetic materials, where the anisotropic spin interactions 
producing topological properties \cite{WK14,Chacon18} for spintronic 
applications drive an increasing focus on first-order QPTs.}

\begin{figure*}
\includegraphics[width=18cm]{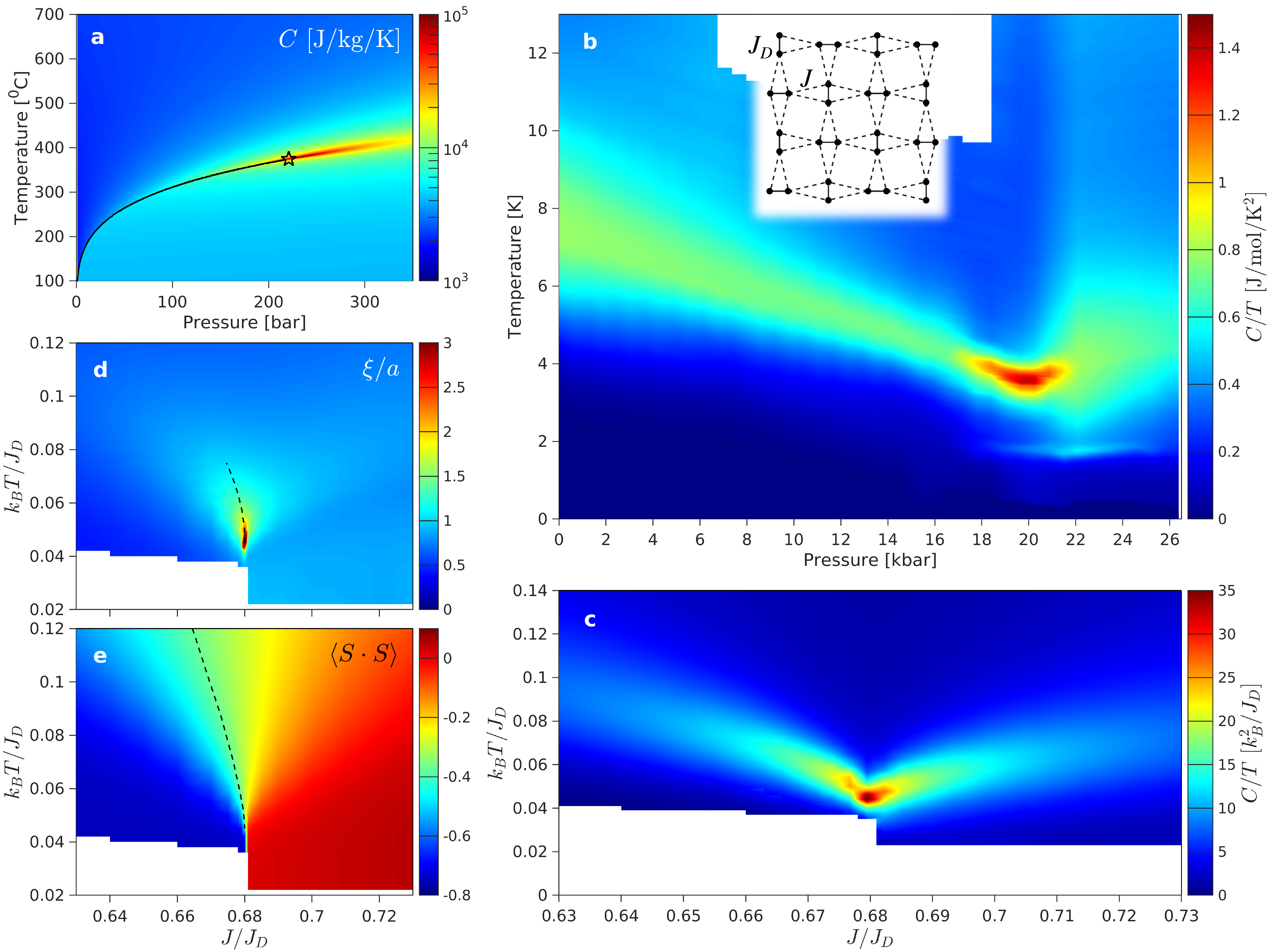}
\caption[]{{\bf Specific heat of water, of SrCu$_2$(BO$_3$)$_2$ and of the 
Shastry-Sutherland model.} {\bf a}, Specific heat of water, $C_p(P,T)/T$, 
shown as a function of pressure and temperature. The black line marks the 
first-order phase boundary separating liquid (lower left) from vapour (upper 
left) and the black star marks the critical point. {\bf b}, Experimental 
data for the specific heat, $C_p(P,T)/T$, of SrCu$_2$(BO$_3$)$_2$. Below 18 
kbar is the dimer product phase and above 20 kbar is the plaquette phase. 
The two lines of maxima meet at the critical point at approximately 
19 kbar and 3.3 K. The inset shows the orthogonal dimer geometry of the 
Shastry-Sutherland model, which is realized by the Cu$^{2+}$ ions ($S = 1/2$) 
in SrCu$_2$(BO$_3$)$_2$. {\bf c}, Analogous data obtained by iPEPS calculations 
with $D = 20$ performed for the Shastry-Sutherland model with different values 
of the coupling ratio, $J/J_D$, which in SrCu$_2$(BO$_3$)$_2$ was shown to be 
an approximately linear function of the applied pressure \cite{Zayed17}. 
{\bf d}, Correlation length, $\xi$, obtained by iPEPS with $D = 20$ and 
expressed in units of the lattice constant, $a$. $\xi$ becomes large only at 
the finite-temperature critical point; the dashed black line shows the locus 
of maxima of $\xi(J/J_D)$ at each fixed temperature, which we terminate when 
$\xi/a < 1$. {\bf e}, Dimer spin-spin correlation function, $\langle {\vec 
S}_i \cdot {\vec S}_j \rangle$, showing a discontinuity with $J/J_D$ at low 
temperatures but continuous behaviour throughout the supercritical regime. 
The dashed black line, the equivalent of the critical isochore in water, 
shows the locus of points where this order parameter is constant at its 
critical-point value, $\langle {\vec S}_i \cdot {\vec S}_j \rangle =
 - 0.372(30)$.}
\label{fig:pd}
\end{figure*}

Water boils at 100$^\circ$C at ambient pressure. At this liquid-gas transition 
the density jumps dramatically, i.e.~the transition is first-order. Under an 
applied pressure, the line of first-order transitions in $(P,T)$ terminates 
at $P_c = 221$ bar and $T_c = 374^\circ$C (Fig.~\ref{fig:pd}a), ``the critical 
point of water,'' where liquid and vapour become a single phase \cite{rcdlt}. 
Because their difference is defined not by a change in symmetry but by a 
scalar, the particle density, which can take two values, the critical point 
is in the Ising universality class. Although the first-order line has no 
critical properties, its termination point (a continuous transition) does 
\cite{rcl}. In the supercritical regime, meaning around and above the critical 
point, there is no transition and indeed one may proceed continuously from 
liquid to vapour without ever undergoing one. 

In quantum matter with active charge degrees of freedom, a critical point 
(which we distinguish from the ``critical endpoint'' \cite{rfn}) terminates 
the Mott metal-insulator transition line \cite{Georges96,Limelette03,Kanoda05}. 
Deep in the insulating phase, systems with only spin degrees of freedom provide 
theorists with a fertile avenue for realizing quantum many-body models, 
including those with exact solutions or ground states, and experimentalists 
with an unparalleled platform for probing their properties on truly macroscopic 
lengthscales. Continuous QPTs and quantum criticality \cite{rsbook} have been 
controlled by pressure \cite{Rueegg08,Merchant14}, applied magnetic field 
\cite{Giamarchi08,Thielemann09}, and disorder \cite{Yu12}. Here we control 
both pressure and field to investigate a discontinuous QPT and provide the 
first observation of critical-point physics in a pure spin system.

\begin{figure*}
\includegraphics[width=16.5cm]{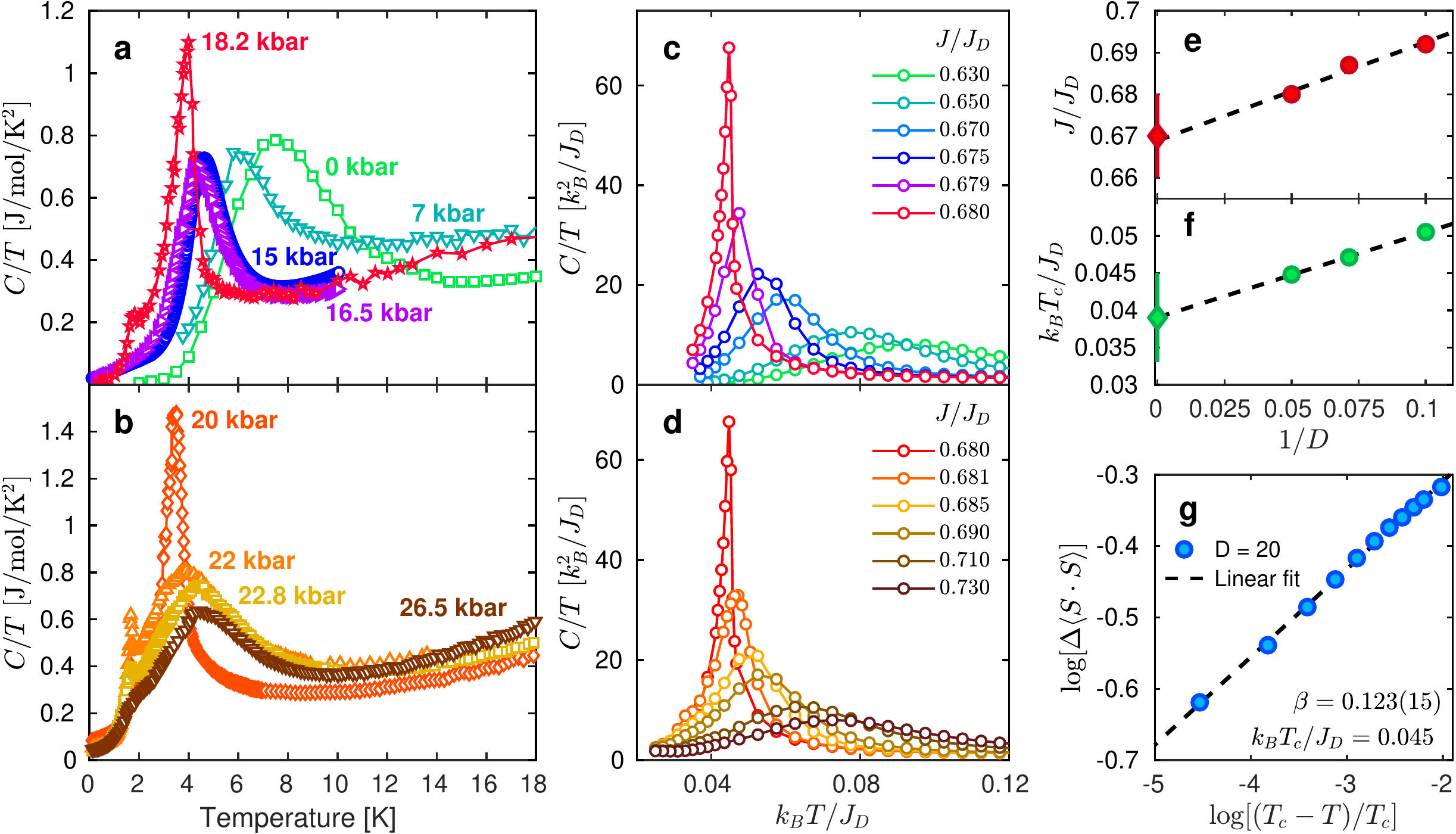}
\caption[]{{\bf Specific heat for different pressures at zero magnetic field.}
{\bf a}, $C_p(T)/T$ measured for applied pressures from 0 to 18.2 kbar and 
{\bf b} from 20 to 26.5 kbar. The temperature of the broad peak observed at 
most pressures ($P \le 15$ kbar and $P \ge 22$ kbar) drops to a finite minimum 
value at the QPT, where $C_p(T)/T$ displays only a tall and extremely narrow 
peak. {\bf c}, {\bf d}, Analogous data obtained by $D = 20$ iPEPS calculations 
performed for the Shastry-Sutherland model in the dimer ({\bf c}) and plaquette 
({\bf d}) phases using plaquette-based tensors (Methods section). The evolution 
of peak heights with proximity to the critical coupling ratio, illustrated 
clearly in the numerical data, is less apparent in experiment. We note that 
the phonon contribution ($C_p(T)/T \propto T^2$) to the measured specific heat 
becomes appreciable at higher temperatures; while this can be subtracted for 
accurate fitting \cite{Wietek19}, our focus here is on the peak positions at 
and below 6 K. {\bf e}, Convergence of the critical coupling ratio obtained in 
finite-temperature iPEPS calculations as a function of $1/D$; the extrapolated 
value of 0.67(1) agrees well with the zero-temperature value \cite{Corboz13}. 
{\bf f}, Convergence of the critical temperature as a function of $1/D$, 
leading to the estimate $k_B T_c/J_D = 0.039(6)$. {\bf g}, Critical exponent, 
$\beta = 0.123(15)$, of the discontinuity, $\Delta \langle {\vec S}_i \cdot 
{\vec S}_j \rangle$, in the dimer spin-spin correlation function, demonstrating 
consistency with the 2D Ising exponent, $\beta = 1/8$.}
\label{fig:CpT}
\end{figure*}

Among the key magnetic quantum materials, SrCu$_2$(BO$_3$)$_2$ \cite{Kageyama99}
has drawn attention because of the extreme frustration of its orthogonal-dimer 
geometry (inset, Fig.~\ref{fig:pd}b). This realizes a $S = 1/2$ Heisenberg 
model formulated by Shastry and Sutherland \cite{ShaSu81} because it has an 
exact ground state, a product of dimer singlets, for all small and intermediate 
interaction ratios ($J/J_D < 0.675$ \cite{Corboz13}). While the frustration is 
manifest in many unusual phenomena \cite{Matsuda13,Knetter00,Wietek19}, our 
interest in the Shastry-Sutherland model lies in the presence of two QPTs, 
from the dimer phase to a plaquette phase at $J/J_D = 0.675(2)$ and thence to 
an ordered N\'eel antiferromagnet (AF) at $J/J_D = 0.765(15)$ \cite{Corboz13}; 
our interest in SrCu$_2$(BO$_3$)$_2$ lies in the remarkable fact that an 
applied hydrostatic pressure acts to control $J/J_D$, revealing both 
transitions at respective pressures of approximately 19 \cite{Zayed17} 
and 27 kbar \cite{Guo20}. 

\begin{figure*}
\includegraphics[width=16.4cm]{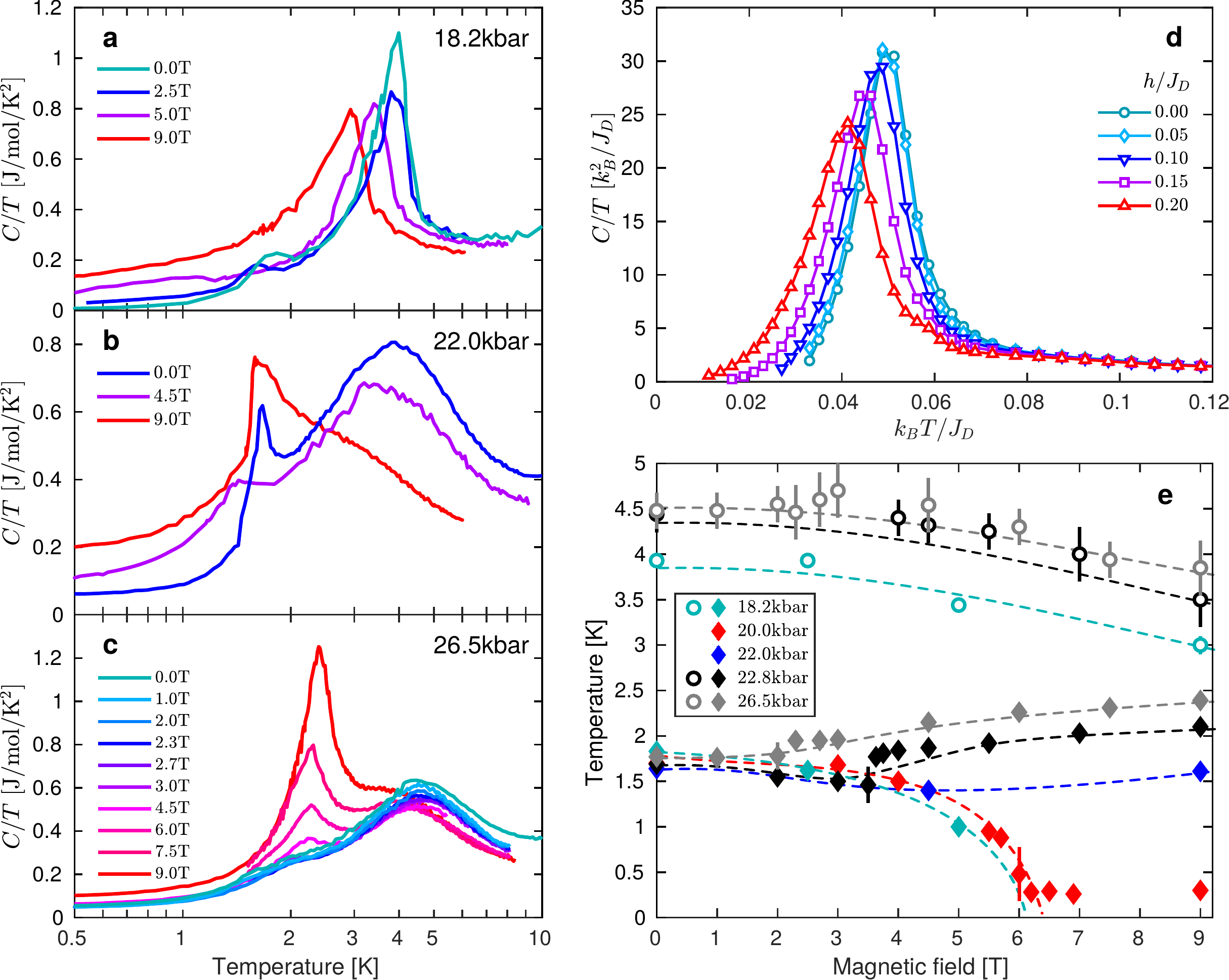}
\caption[]{{\bf Evolution of the specific heat with magnetic field for 
pressures around the QPT.} {\bf a}, $C_p(T)/T$ at 18.2 kbar for four different 
applied fields up to $\mu_0 H = 9$ T. The peak due to the critical point 
retains its sharp nature and moves only slightly downwards in position and 
height with increasing field. {\bf b}, $C_p(T)/T$ at 22.0 kbar for three 
different applied fields, showing initial field-induced suppression of the 
low-temperature transition followed by a dramatic change in shape to a sharp 
low-$T$ peak with no broad hump at higher energies. {\bf c}, $C_p(T)/T$ at 
26.5 kbar, illustrating the field-induced emergence of the new, sharp peak at 
9 T, which is suggestive of the transition to the AF phase. {\bf d}, $C_p(T)/T$ 
computed by iPEPS at a coupling ratio, $J/J_D = 0.686$, close to the critical 
point for $D = 14$, for a range of applied magnetic fields. As in panel {\bf a},
the sharp peak due to the critical point undergoes only a minor field-induced 
suppression of its position and height, while its width is unaltered. For the 
$g$-factor of SrCu$_2$(BO$_3$)$_2$, a field $h/J_D \equiv g \mu_B \mu_0 H 
/J_D = 0.2$ corresponds to approximately 11 T. {\bf e}, Characteristic field 
and temperature scales revealed by our full set of specific-heat measurements, 
which separate into high-$T$ features around 4 K and low-$T$ features around 
and below 2 K. The high-$T$ feature is the broad maximum, which at the critical 
point evolves into the sharp peak. The low-$T$ feature at the lower pressures 
is the small peak due to the thermal Ising transition out of the plaquette 
phase, which can be suppressed to $T = 0$ by the applied field. At the higher 
pressures this feature changes into the strong peaks arising from the thermal 
transition of the AF phase, which is favoured by the applied field.}
\label{fig:CpB}
\end{figure*}

The finite-temperature critical point has been discussed theoretically 
in a two-dimensional (2D) Heisenberg spin model, the ``fully frustrated 
bilayer'' \cite{Stapmanns18}. In this geometry, which has no known materials 
analogue, spin pairs (with coupling $J_\perp$) are arranged vertically on a 
square lattice with equal couplings ($J_\|$) to both spins of all four dimer 
neighbours, and the ground state jumps discontinuously from exact dimer 
singlets to exact triplets at $J_\perp/J_\| = 2.315$. With increasing 
temperature, the discontinuity in triplet density reduces until the line 
of first-order transitions terminates at a critical point, in the 2D Ising 
universality class, when $k_B T_c \simeq 0.52 J_\|$ \cite{Stapmanns18}. The 
connection \cite{Wessel18} between the two fully frustrated geometries 
(bilayer and Shastry-Sutherland) suggests that they may share similar 
critical-point physics. Although the total spin of each dimer is not a good 
quantum number in the Shastry-Sutherland case, and hence it may not have the 
same large $T_c$ as the bilayer, a discontinuity in the average dimer spin-spin 
correlation is already well known at the dimer-plaquette QPT. 

To investigate the critical point in SrCu$_2$(BO$_3$)$_2$, we perform 
high-precision measurements of the specific heat using an a.c.~calorimetry 
technique \cite{Larrea20}. As described in the Methods section, large single 
crystals of SrCu$_2$(BO$_3$)$_2$ were grown by a floating-zone method. Samples 
of masses up to 36 mg were cut, patterned with metallic strips for calorimetry 
and mounted in a clamp cell allowing hydrostatic pressures up to 26.5 kbar and 
in applied magnetic fields up to 9 T. Details of our a.c.~measurement 
procedures are provided in the Methods section. 

Starting with zero magnetic field, the pressure-induced evolution of the 
specific heat, shown as $C_p(T)/T$, is illustrated in Fig.~\ref{fig:pd}b. 
As quantified in Fig.~\ref{fig:CpT}a, $C_p(T)/T$ at low pressures shows an 
exponential rise to a broad maximum at a temperature, $T_{\rm max}$, that 
tracks the gap to the triplon or bound-triplon excitations of the dimer 
phase \cite{Wietek19}. With increasing $P$, this peak moves gradually lower 
and becomes proportionately narrower, but between 18 and 20 kbar it becomes 
extremely narrow and attains a much higher peak value (Figs.~\ref{fig:pd}b 
and \ref{fig:CpT}a), bearing all the characteristics of a continuous phase 
transition, with diverging observables. After reaching a lowest measured 
value of 3.4 K at $P = 20$ kbar, $T_{\rm max}$ rises with increasing pressure 
and the peak broadens again (Fig.~\ref{fig:CpT}b), indicating that the 
singular behaviour has terminated. 

A second small peak appears around 2 K for $P \ge 18$ kbar (Figs.~\ref{fig:pd}b 
and \ref{fig:CpT}a) and persists to our upper pressure limit. We expect that 
this peak corresponds to the thermal transition out of the plaquette phase and 
provide a detailed discussion below. Our key observation is that the critical 
point occurs at a temperature well above the plaquette transition, consistent 
with its interpretation as the termination of a line of first-order transitions.

To model these thermodynamic results we use the method of infinite projected 
entangled pair states (iPEPS), which are a variational tensor-network Ansatz
for the representation of a quantum state on an infinite lattice 
\cite{Verstraete04,Jordan08}. The accuracy of the Ansatz is controlled by 
the bond dimension, $D$, of the tensors (Methods section). While iPEPS have 
been applied previously to discuss the ground state of the Shastry-Sutherland 
model \cite{Corboz13,Matsuda13,Boos19}, here we apply newly-developed methods 
\cite{Czarnik19,Wietek19} for representing the thermal states of the system. 
We analyse the pure Shastry-Sutherland model, meaning a single 2D layer 
(inset, Fig.~\ref{fig:pd}a); SrCu$_2$(BO$_3$)$_2$ is known to have 
weak and frustrated inter-layer interactions (at most 10\% of $J_D$), 
Dzyaloshinskii-Moriya (DM) interactions (3\% \cite{Nojiri99}) and higher-order 
further-neighbour interactions, none of which affect the first-order nature of 
the dimer-plaquette transition.

iPEPS results for $C_p(T)/T$ as a function of $J/J_D$, presented in 
Fig.~\ref{fig:pd}c, show the same evolution as in SrCu$_2$(BO$_3$)$_2$ under 
pressure. The broad peaks of the gapped dimer and plaquette states move to 
lower temperatures on approaching the QPT and narrow to a tall, sharp spike 
at the critical coupling ratio (Figs.~\ref{fig:CpT}c,d). Figure \ref{fig:pd}d 
shows that the correlation length grows dramatically around the critical point, 
but remains small at temperatures below it. Figure \ref{fig:pd}e illustrates 
how the average dimer spin-spin correlation, a scalar which serves as an order 
parameter for the nature of the spin state, jumps discontinuously at $T < T_c$ 
but changes to a smooth function of $J/J_D$ at $T \ge T_c$, in direct analogy 
to the density of water molecules.

Our iPEPS results computed with $D = 10$, 14 and 20 show the same qualitative 
forms and provide good quantitative convergence (Figs.~\ref{fig:CpT}e,f) 
towards the critical coupling ratio of zero-temperature iPEPS \cite{Corboz13}. 
From this we estimate the critical temperature $k_B T_c = 0.039(6) J_D$. We 
work in units of $J_D$ because the exact $P$-dependence of the magnetic 
interactions in SrCu$_2$(BO$_3$)$_2$ is subject to further uncertainty; taking 
a linear extrapolation at constant $J$, with errors provided by alternative 
estimates \cite{Zayed17}, a value $J_D (P_c)/k_{\rm B} = 77(8)$ K gives 
a best estimate of $T_c = 3.0(6)$ K, in excellent agreement with experiment. 
By analysing the critical scaling of the discontinuity for $D = 20$, we deduce 
(Fig.~\ref{fig:CpT}g) that the exponent is fully consistent with the value 
$\beta = 1/8$ expected of a 2D Ising transition. iPEPS calculations in the 
critical regime become increasingly challenging at lower temperatures 
(Methods), and a numerical instability occurs in the dimer-phase regions 
excluded from Figs.~\ref{fig:pd}c-e and \ref{fig:CpT}c-d. However, these 
regions are readily accessed by a different iPEPS Ansatz \cite{Wietek19}, 
and also by working with a finite DM interaction, as shown in Extended Data 
Fig.~ED2. This confirms further that the DM interactions of SrCu$_2$(BO$_3$)$_2$
have no effect on the physics of the critical point. 


To challenge our interpretation of the critical point, we consider the 
situation in a finite magnetic field. Because the field has little effect 
on the gapped dimer and plaquette phases, the physics of the critical point 
should be essentially unaffected. We have measured the specific heat in fields 
up to $\mu_0 H = 9$ T, and indeed we observe near the critical point (18.2 
kbar, Fig.~\ref{fig:CpB}a) that the peak remains sharp and shows only very 
minor field-induced changes. These features are reproduced in detail by our 
iPEPS results, shown in Fig.~\ref{fig:CpB}d for a $J/J_D$ ratio very close to 
the $D = 14$ QPT, which also indicate that the small changes can result simply 
from not pinpointing the exact critical coupling. 

\begin{SCfigure*}
\includegraphics[width=11cm]{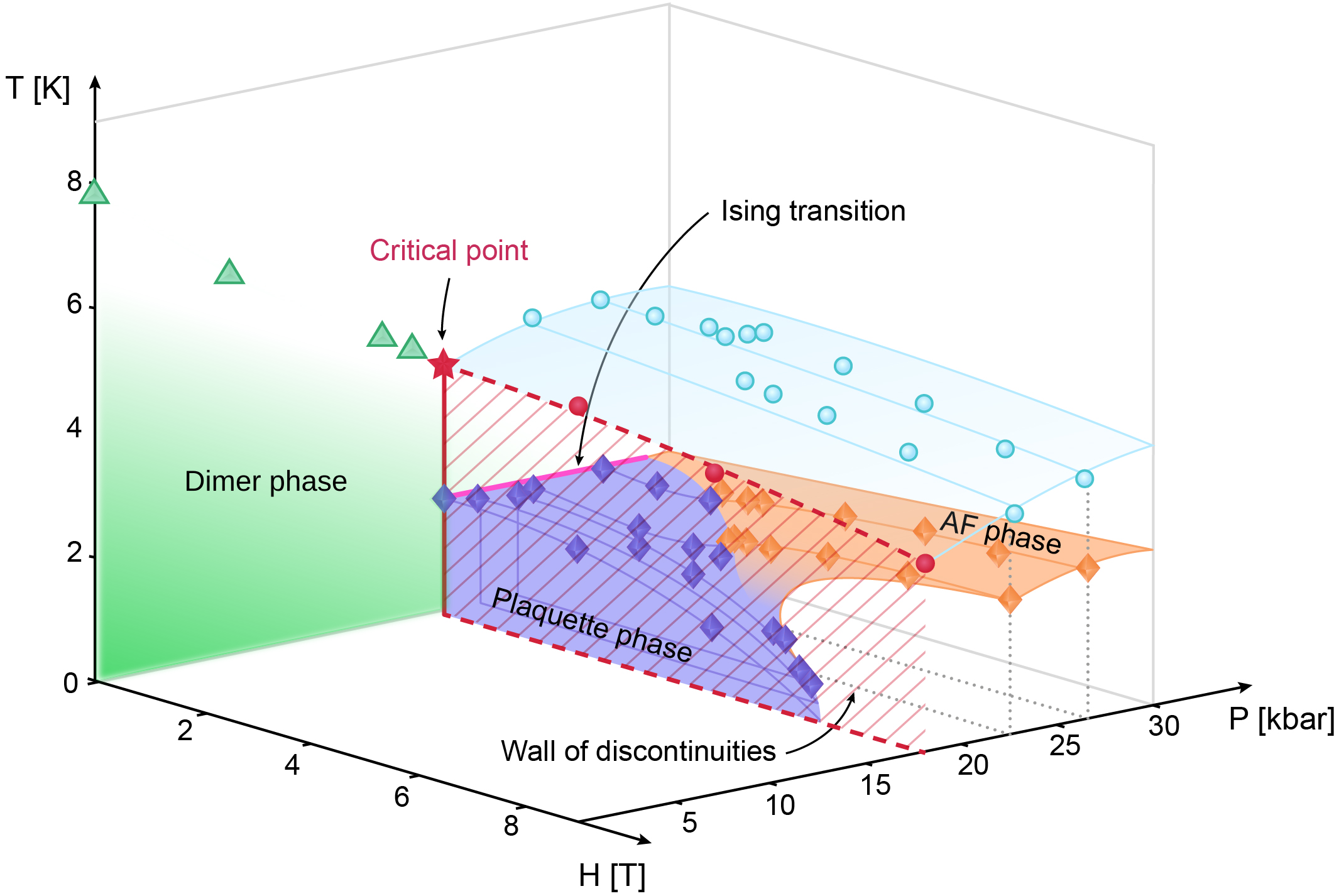}
\caption[]{{\bf Schematic phase diagram in field, pressure and temperature.}
Phase diagram of SrCu$_2$(BO$_3$)$_2$ illustrated in the full $(P,H,T)$ 
space. The turquoise surface and green triangles indicate the locations 
of the maxima in $C_p(T)/T$, respectively in the plaquette and dimer phases.
These evolve into the critical point (red star, $T_c = 3.3$ K) in the $(P,T)$ 
plane. The solid vertical red line below the critical point shows the location 
of the line of first-order discontinuities at $H = 0$; the dashed red lines 
and red grading mark the wall of discontinuities that extends to finite 
magnetic field and terminates at critical temperatures (red circles) that 
fall only slowly with $H$. The horizontal magenta line marks the measured 
location of the Ising transition out of the plaquette phase at $H = 0$. The 
blue surface marks the volume of parameter space occupied by the plaquette 
phase: its upper surface is the continuation of the magenta transition line 
to finite fields and at low pressures it is delimited by the wall of 
discontinuities. The orange colour indicates the surface of thermal 
transitions out of the AF phase.} 
\label{fig:pd_3D}
\end{SCfigure*}

Turning to the 2 K peak, our 18.2 and 22.0 kbar data (Figs.~\ref{fig:CpB}a,b) 
show that the field causes a strong suppression of both its height and 
position. However, at 9 T for 22.0 kbar, another peak has emerged that is 
quite different in shape, becoming tall, sharp and isolated from the broad 
4 K hump, which has almost vanished. This behaviour is also observed above 
4 T in our 26.5 kbar data (Fig.~\ref{fig:CpB}c). In Fig.~\ref{fig:CpB}e we 
collect all of these peaks to obtain a clear picture of three key phenomena. 
First, the 4 K features remain isolated, changing from sharp to broad with 
increasing $P$ and $H$ quite independently of the complex action below 2 K. 
Second, the thermal transition of the plaquette phase, occurring at 2 K at 18.2 
kbar, is suppressed to zero by fields of 6 T, even at higher pressures that 
are expected to stabilize it. Third, the new, sharp peak appears to emerge at 
finite temperature at higher pressures and fields; we believe this to be the 
ordering transition of the AF phase, but cannot state definitively whether it 
is also first-order. Because of the challenges encountered by our iPEPS 
calculations at very low temperatures, we are not currently able to model 
these transitions reliably. 

Here we comment that similar $C_p(T)$ data in zero field have appeared recently 
\cite{Guo20}. These authors did observe the tall, sharp critical-point peak in 
their 1.9 and 2.1 GPa data (shown only in their Supplemental Materials), but 
provide no explanation. They also observe the 2 K peak and ascribe it to a 
plaquette state (which they assume to be the EP phase, below). At pressures 
beyond the limits of our study, they argue that the 2 K feature develops into 
a signature of the AF phase, but do not find the strong, sharp features we 
obtain in an applied field (Figs.~\ref{fig:CpB}b,c).


Turning to the theoretical underpinnings of the critical-point phase diagram, 
NMR \cite{Waki07} and neutron scattering experiments \cite{Zayed17} indicate 
that the low-temperature plaquette phase of SrCu$_2$(BO$_3$)$_2$ is not that 
of the Shastry-Sutherland model, where singlets form in the ``empty'' 
plaquettes (EP) of the $J$ lattice, meaning those with no $J_D$ dimer 
(inset, Fig.~\ref{fig:pd}b), but in the ``full'' plaquettes (FP). 
From the sensitive energetic competition between the EP and FP phases 
\cite{Boos19}, it is not surprising that the additional 3D and DM terms 
in SrCu$_2$(BO$_3$)$_2$ could cause this discrepancy, so we expect that the 
state we observe above 18 kbar and below 2 K is the FP phase. The important 
property of both plaquette phases is their two-fold degenerate ground state 
(only half of the plaquettes may form singlets), and thus the thermal 
transition out of the plaquette phase has Ising symmetry. The other important 
observation is that this transition occurs at 2 K, well below the critical 
point. 

This Ising character contrasts with the fully frustrated bilayer, where the 
QPT is straight to a 2D AF phase that by the Mermin-Wagner theorem has no 
finite-temperature transition, implying that the critical point associated 
with the discontinuity can only be observed as an isolated point, as in 
water. In the Shastry-Sutherland model, where the plaquette phase breaks 
a Z$_2$ symmetry, there are {\it a priori} two possibilities for connecting 
the Ising transition line to the first-order transition line extending from 
the QPT to the critical point. (i) The Ising transition line ends when 
it touches the first-order line at a temperature $T < T_c$, forming a critical 
endpoint \cite{rfu,rfb,Stapmanns18}. (ii) The point $(P_c,T_c)$ is in fact a 
tricritical Ising point, where the first-order line turns continuously into 
the Ising line. Possibility (i) is clearly favoured by our specific-heat 
data, which show the critical point at $P_c \simeq 19$ kbar and $T_c \simeq 
3.3$ K to be well isolated and to exhibit no anomalies around $T_c$, while 
the intrinsic critical temperature of the plaquette phase remains below 2~K.

We represent this situation in the ($P,T$) plane of Fig.~\ref{fig:pd_3D}. On 
the scale of the figure, the first-order line is essentially vertical in 
temperature \cite{Stapmanns18}. Although a structural transition is probably 
associated with the magnetic transition in SrCu$_2$(BO$_3$)$_2$, causing a 
discontinuity in the coupling constants, we stress that there is only one 
transition as a function of $J/J_D$ (Fig.~\ref{fig:pd}). In Fig.~\ref{fig:pd_3D}
we have gathered all the features abstracted from our data in field as well as 
pressure. We find that the picture of the critical point at $H = 0$ is 
supported by the data at finite fields, which indicate a line of critical 
points, and a wall of first-order transitions, that dominate the ($P,H,T$) 
phase diagram. Below $T_c(P,H)$, this wall defines a plane through which the 
average dimer spin-spin correlation should change discontinuously, a prediction 
that may be tested by measuring the $P$-dependence of the instantaneous spin 
correlation function by neutron scattering, or of phonon modes sensitive to 
the magnetic correlations \cite{Bettler20}.

Below the critical point in Fig.~\ref{fig:pd_3D}, our data reveal a 
wealth of quantum and thermal phase transitions occurring between 0 and 2 K. 
The field suppresses the continuous thermal transitions out of the plaquette 
phase and in its place we have found the AF phase, for the first time at 
pressures as low as 22.0 kbar. The pressure-induced plaquette-AF QPT is 
thought to be weakly first-order \cite{Corboz13}, although it has recently 
been proposed as a candidate deconfined quantum critical point \cite{Lee19}. 
While our data in different regimes suggest that the field-induced plaquette-AF 
QPT could be first- or second-order, there is no clear sign of a second 
finite-temperature critical point. An experimental verification of the 
universality classes of these transitions poses a challenge to specific-heat 
measurements under such extreme conditions of pressure, field and temperature. 
Nevertheless, given the recent numerical and experimental progress in 
probing the thermal properties of frustrated systems, such verification may 
soon become possible. 

The physics of the critical-point phase diagram appears rather different 
from that of second-order QPTs: because one may pass from one side of the 
discontinuity to the other without crossing a transition, there is no breaking 
of symmetry (here SU(2) spin). At the critical point itself, the property of a 
divergent correlation length (Fig.~\ref{fig:pd}d), associated with domain sizes 
and having critical exponents set by the universality class, remains. The 
supercritical regime has recently become a subject of active investigation, 
even (with a view to sensitive switching) in critical fluids \cite{Maxim19}. 
In the Mott metal-insulator phase diagram, theory \cite{Terletska11,Vojta19} 
and experiment \cite{Kanoda15} have suggested the emergence of quantum critical 
scaling in this regime. In SrCu$_2$(BO$_3$)$_2$, the striking feature of the 
phase diagram is that the temperature, $T_{\rm max}$, characterizing the peak in 
$C_p/T$ reaches a minimum at $T_c$. We show in Fig.~ED3 that this behaviour 
is universal around an Ising critical point in 2D lattice models. Thus the 
specific heat clearly defines not one but two characteristic lines in the 
supercritical regime (Figs.~\ref{fig:pd}b-c), in contrast to the correlation 
length (Fig.~\ref{fig:pd}d) and the critical isochore (Fig.~\ref{fig:pd}e), 
both of which are regarded as marking a single crossover line. Although 
this remarkable property of the specific heat is quite different from 
water (Fig.~\ref{fig:pd}a), we stress that it is intrinsic to a model as 
simple as the Ising model. While the origin of this complex physics deserves 
further theoretical analysis, we observe that both types of behaviour can be 
probed experimentally in quantum spin systems by comparing the specific heat 
with scattering measurements of the order parameter. 

In summary, we have shown that the first-order QPT in the quantum magnetic 
material SrCu$_2$(BO$_3$)$_2$ is accompanied by a finite-temperature critical 
point analogous to the phase diagram of water. Our studies span a wide range 
of pressures and applied magnetic fields around the critical regime and, by 
revealing the connectivity of the phase diagram and the location of the 
antiferromagnetic phase, illustrate the importance of controlling the 
environment to complete studies of criticality and universality. We have 
explained our experimental data by powerful new numerical methods allowing 
access to the finite-temperature regime of frustrated systems in two 
dimensions. As modern quantum magnetism and spintronics embrace spin-orbit 
coupling and the resulting highly spin-anisotropic interactions required to 
produce the topological physics of Ising, Kitaev, skyrmion and other systems, 
a full understanding of the resulting first-order QPTs will include the 
quantum phenomenology of the critical point, whose classical variant has 
been known to science for two centuries. 

\bibliographystyle{naturemag}
\bibliography{cp}

\begin{thebibliography}{10}
\bibitem{rfn}
\bibinfo{title}{{Because the first-order line has no critical properties, only
  its termination point is a ``critical point.'' Following M. E. Fisher
  \cite{rfu,rfb} we reserve the terminology ``critical endpoint'' for the
  termination of a line of continuous (second-order) phase transitions
  \cite{Stapmanns18}}} .
\end{thebibliography}

\bigskip
\noindent
{\bf Acknowledgements}

\noindent
We are grateful to R. Gaal, J. Piatek and M. de Vries for technical assistance. 
We acknowledge helpful discussions with D. Badrtdinov, C. Boos, T. Fennell, A. 
Turrini, A. Wietek and A. Zheludev. 
We thank the S\~{a}o Paulo Research Foundation (FAPESP) for financial support 
under Grant No.~2018/08845-3, the Qatar Foundation for support through 
Carnegie Mellon University in Qatar's Seed Research programme, the Swiss 
National Science Foundation (SNSF) for support under Grant No.~188648 and 
the European Research Council (ERC) for support under the EU Horizon 2020 
research and innovation programme (Grant No.~677061), as well as from the 
ERC Synergy Grant HERO. We are grateful to the Deutsche Forschungsgemeinschaft 
for the support of RTG 1995 and to the IT Center at RWTH Aachen University and 
the JSC J\"ulich for access to computing time through JARA-HPC.

\bigskip
\noindent
{\bf Author contributions}

\noindent
The experimental project was conceived by H.M.R. and Ch.R. and the theoretical 
framework was put forward by F.M. The crystals were grown by E.P. and K.C. 
Specific-heat measurements were performed by J.L.J. with assistance from 
M.E.Z., R.L. and H.M.R. S.C. and P.C. performed iPEPS calculations. A.L. 
performed complementary exact diagonalization calculations. L.W. and S.W. 
performed quantum Monte Carlo calculations on the fully frustrated bilayer 
model. Data analysis and figure preparation were performed by J.L.J., E.F., 
S.C., L.W., S.W., P.C. and H.M.R. The detailed theoretical analysis was 
provided by P.C., S.C., F.M., A.H., B.N., L.W. and S.W. The manuscript was 
written by B.N. and F.M. with assistance from all the authors.

\bigskip
\noindent
{\bf Additional information}

\noindent
The authors declare no competing financial interests. The statements made 
herein are not the responsibility of the Qatar Foundation. Correspondence 
and requests for materials should be addressed to H.M.R. 
(henrik.ronnow@epfl.ch).

\setcounter{figure}{0}
\renewcommand{\thefigure}{ED\arabic{figure}}

\setcounter{equation}{0}
\renewcommand{\theequation}{S\arabic{equation}}

\setcounter{table}{0}
\renewcommand{\thetable}{S\arabic{table}}

\onecolumngrid

\vskip10mm

\noindent
{\large {\bf {Methods Section and Extended Data Figures for}}}

\vskip4mm

\noindent
{\large {\bf {A quantum magnetic analogue to the critical point of water}}}

\vskip4mm

\noindent
J. Larrea Jim\'enez, S. P. G. Crone, E. Fogh, M. E. Zayed, R. Lortz, 
E. Pomjakushina, K. Conder, A. M. L\"auchli, L. Weber, S. Wessel, 
A. Honecker, B. Normand, Ch. R\"uegg, P. Corboz, H. M. R{\o}nnow and F. Mila, 

\vskip4mm

\noindent
{\large {\bf Methods}}

\vskip2mm

\twocolumngrid

\noindent
{\bf Samples.} Single crystals of SrCu$_2$(BO$_3$)$_2$ were grown by a 
floating-zone method. First polycrystalline SrCu$_2$(BO$_3$)$_2$ was prepared 
by a solid-state reaction using as starting materials SrCO$_3$, CuO and 
B$_2$O$_3$ with 99.99\% purity. These were mixed, ground and heat-treated at 
900$^\circ$C in flowing oxygen for over 100 h with several intermediate 
grindings. The phase purity was verified by conventional x-ray diffractometry 
and the powder was pressed hydrostatically into rods (8 mm in diameter and 
around 90 mm in length) which were sintered at 900$^\circ$C for 20 h in an 
oxygen atmosphere.

Single-crystal growth was carried out using an Optical Floating Zone Furnace 
(FZ-T-10000-H-IV-VP-PC, Crystal System Corp., Japan) with four 300 W halogen 
lamps as the heat source. Although the first reported single crystals of 
SrCu$_2$(BO$_3$)$_2$ were also grown in this type of furnace, using LiBO$_2$ 
discs as a solvent placed on a top of a seed rod \cite{Kageyama99c}, we
obtained better results using a self-adjusted flux method \cite{Jorge04}. 
Optimal growth conditions were found to be a steady growth rate of 0.25 mm/h 
at all times, rotation of both feeding and seeding rods at approximately 15 
rpm in opposite directions to ensure homogeneity of the liquid and the 
application of a 5 bar pressure of argon with 20\% oxygen. When a homogeneous 
melt was achieved, the power of the lamps was decreased slowly (over 24 hours) 
until steady-state conditions were found (c.~48.5\% lamp power in the present 
case), after which these conditions were maintained rigorously until the end 
of the growth. The full growth process required approximately two weeks and 
resulted in a recrystallized boule of 8 cm in length containing a 
single-crystalline grain in its top 4 cm. 

\medskip
\noindent
{\bf Specific-heat measurements.} 
Three slabs were cut from the single-crystal rod, each with a cross-section 
of 7.6-9.0 mm$^2$ in the $ab$ plane and a thickness of 0.5-1.0 mm parallel
to the $c$-axis. The sample masses were measured to be between 15 and 36 mg. 
The samples were polished gently to deposit Pt films of 20 nm thickness.
To measure the heat capacity, $C(P, H, T)$, under multiple extreme conditions 
(very low temperatures, high pressures and intense magnetic fields), we used 
alternating-current (a.c.) calorimetry at a second-harmonic mode 
(``2$\omega$''). We employed a piston cylinder BeCu clamp cell to apply 
pressures up to 26.5 kbar, with liquid kerosene as the pressure-transmitting 
medium. Measurements of the electrical resistivity at the superconducting 
transition of a Pb strip were used as a pressure manometer and revealed good 
hydrostatic conditions in our experiments. The pressure cell was inserted into 
a cryostat with a $^3$He-$^4$He dilution refrigerator that allowed both cell 
and sample to be cooled to 0.1 K with temperature variations below 1 mK. At 
some pressure steps, a magnetic field up to 9 T was applied using a 
superconducting magnet whose field direction was perpendicular to the $c$-axis.

\begin{figure*}
\hspace{0.4cm}\includegraphics[width=7.7cm]{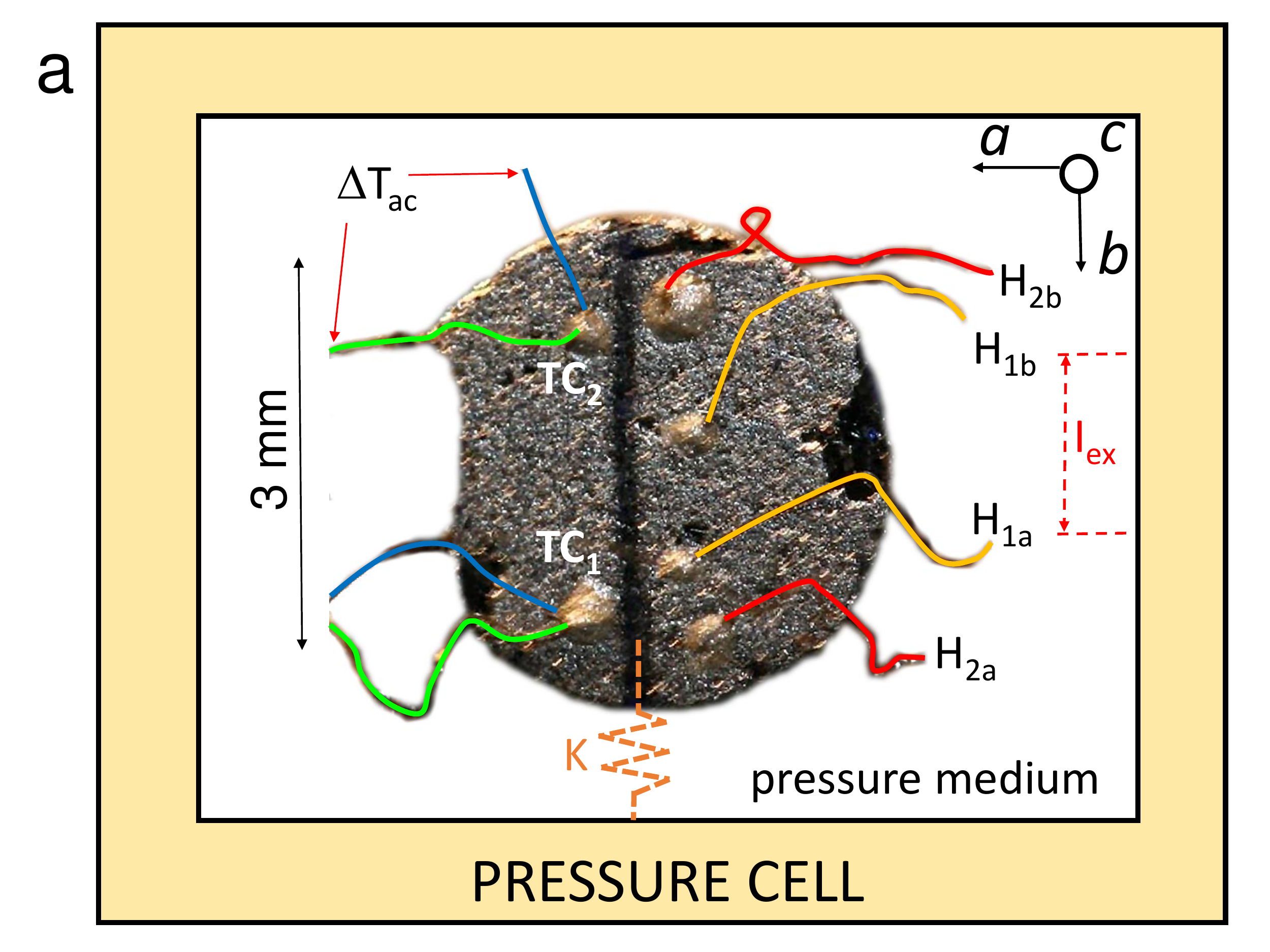}\hspace{0.4cm}
\includegraphics[width=8.5cm]{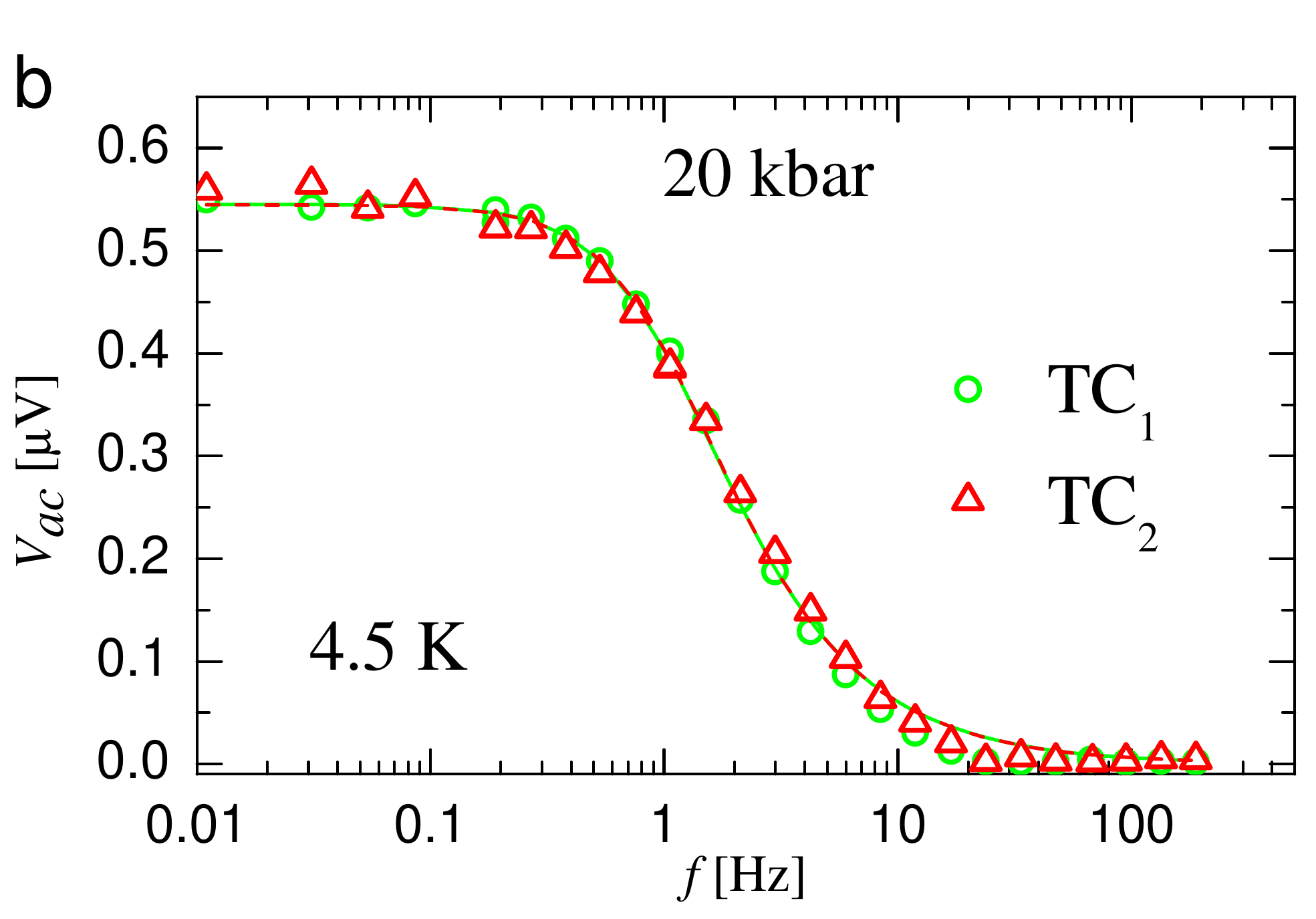}
\includegraphics[width=8.5cm]{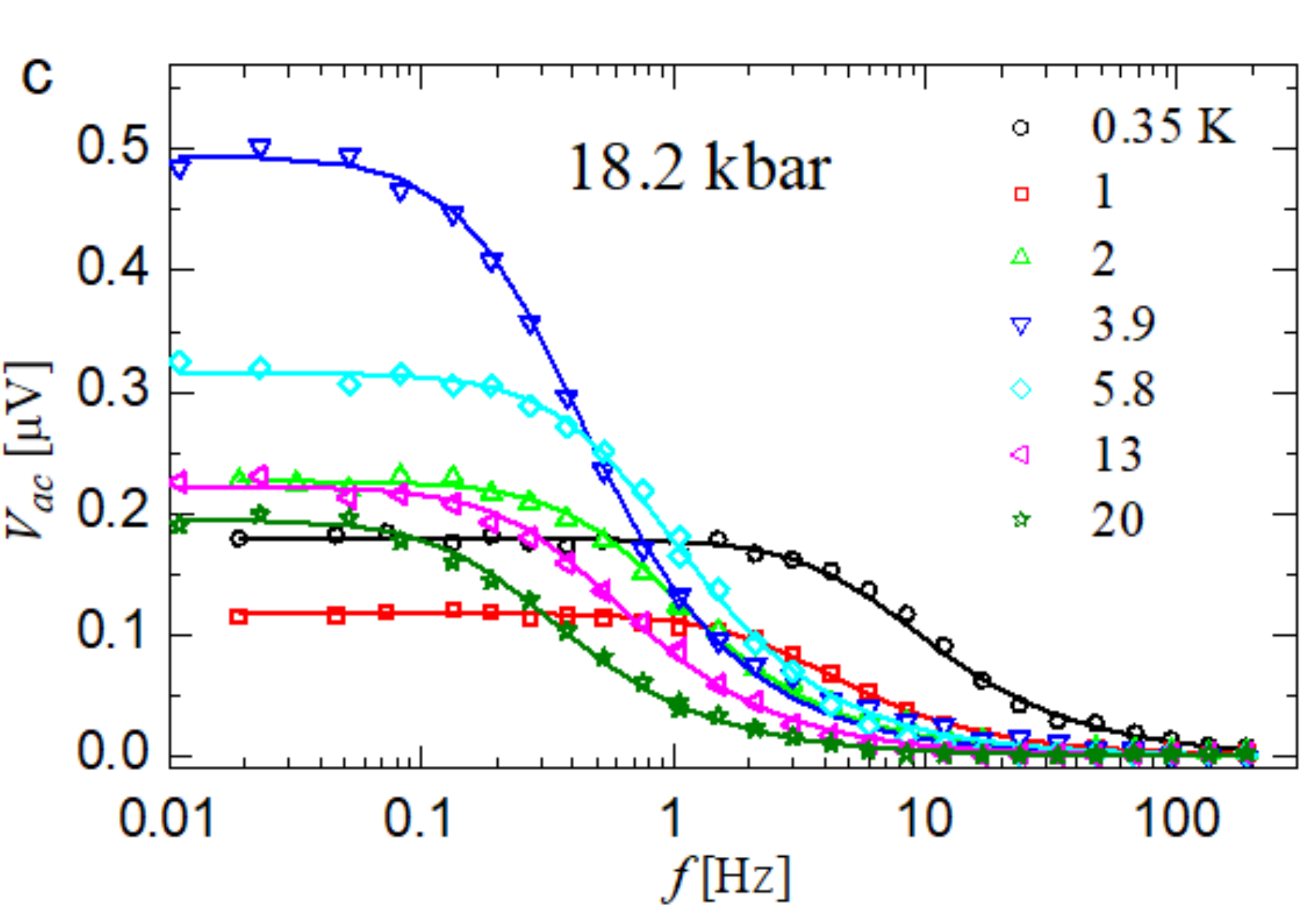}
\includegraphics[width=8.5cm]{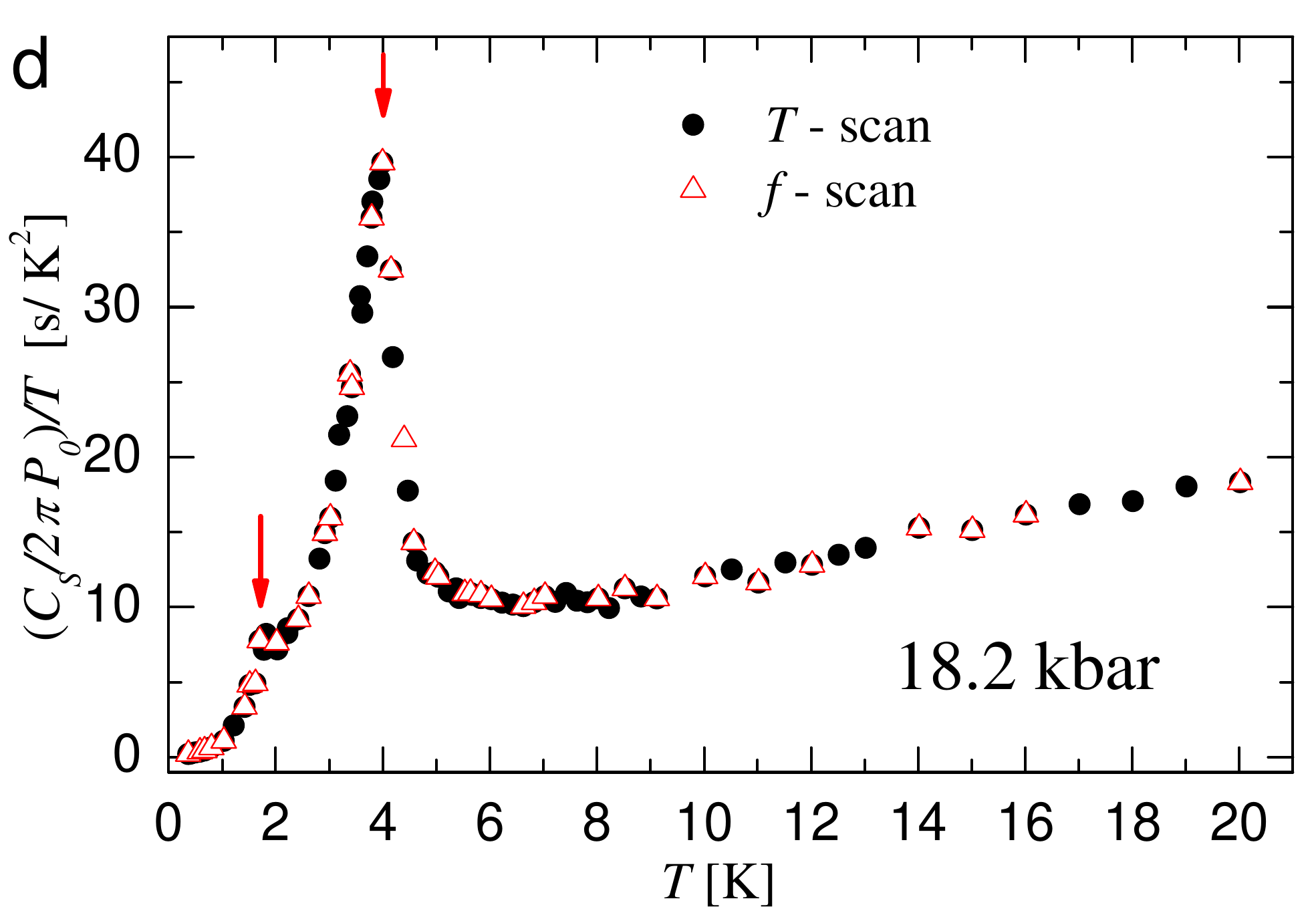}
\caption[]{{\bf a.c.~calorimetry on SrCu$_2$(BO$_3$)$_2$}.
{\bf a}, The a.c.~calorimeter was prepared by depositing two Pt thin films 
(the shinier surfaces) over both halves of the sample. One film was used as 
the heater and the other for optimal thermal contact and measurement. The 
heating current ($I_{\rm ex}$ at frequency $f$) was supplied through the pair of 
Constantan wires, H$_{\rm 1a}$ and H$_{\rm 1b}$, while H$_{\rm 2a}$ and H$_{\rm 2b}$ 
were used to measure the electrical resistance, $R_{\rm Pt}$, of the Pt film. 
TC$_1$ and TC$_2$ are thermocouples and $K$ represents the thermal contact 
between the sample and the cryostat (through the pressure cell). 
{\bf b}, Isothermal ($T = 4.5$ K) and isobaric ($P = 20$ kbar) $f$-dependence 
of the modulated pick-up voltage, $V_{ac}$, which is directly proportional to 
the temperature differential, $\Delta T_{ac}$, measured by the thermocouples 
at two different positions.
{\bf c}, Isobaric ($P = 18.2$ kbar) $f$-dependence measurements of $V_{ac}$ at 
different temperatures with $I_0 = 1.6$ mA at $T \ge 3.9$ K, $I_0 = 0.8$ mA at 
$T = 2$ K and $I_0 = 0.4$ mA at $T < 2$ K. 
{\bf d}, Sample heat capacity normalized to the input heating power ($P_0
 = I_0^2 R_{\rm Pt}$), comparing the fit of $V_{ac} (f)$ obtained from the 
steady-state equation (``$f$-scan'' \cite{Larrea20}) with values obtained 
directly from a variable-temperature measurement performed at the fixed 
working frequency $f_C = 1.5$ Hz (``$T$-scan''). Solid and dashed lines in 
panels {\bf b} and {\bf c} represent fits using the steady-state equation 
\cite{Larrea20,Gmelin97}. }
\label{facc}
\end{figure*}

Our a.c.~calorimetry set-up was sputtered onto the sample slab to guarantee  
good thermal contact of the sample, heater and thermometer \cite{Larrea20}. 
It consisted of two separate Pt films, one used as the heater and the other 
used to improve the thermal contact between sample and thermometer, as shown 
in Fig.~\ref{facc}a. This configuration allowed the efficient measurement of
the modulated temperature difference in the sample ($\Delta T_{ac}$) as a 
function of the amplitude and frequency, $f = \omega/2\pi$, of the alternating 
excitation current, $I_{\rm ex} = I_0 e^{-i \omega t}$. The heating power was applied
through two Constantan wires (H$_{\rm 1a}$ and H$_{\rm 1b}$ in Fig.~\ref{facc}a), 
this choice of metal being made to avoid heat leakage through the wires; the 
gold wires (H$_{\rm 2a}$ and H$_{\rm 2b}$) were applied to measure the electrical 
resistance of the heater as a function of $P$, $H$ and $T$. We used a 
AuFe(0.07)/Chromel thermocouple (TC$_2$ in Fig.~\ref{facc}a) as the 
thermometer detecting the temperature differential, $\Delta T_{ac}$, 
between the sample and its environment, and placed a second thermocouple 
(TC$_1$) symmetrically opposite to it; because TC$_1$ showed the same 
$f$-dependence of $\Delta T_{ac}$ as TC$_2$, we could conclude that there 
were no thermal gradients across the sample within the resolution of our 
experiment.

The sample heat capacity was obtained directly from the isothermal $f$-scans, 
shown in Figs.~\ref{facc}b-c, by fitting with the standard steady-state 
equation \cite{Gmelin97}. Given the complete control of all relevant 
a.c.~calorimetry parameters, our method has a number of advantages in the 
determination of absolute values of $C(P, H, T)$ when compared with previous 
work \cite{Guo20}. These advantages include a more precise determination of 
the corrections for unwanted heat losses throughout the sample, a more precise 
separation of additional $P$-dependent contributions and the ability to achieve 
an optimal thermal equilibrium inside the sample (within the resolution of the 
thermocouple). Thus, our ``$f$-scan'' analysis \cite{Larrea20} allows an 
accurate determination of the working frequency ($f_C$) at which to maximize 
$C(P, H, T)$ at each pressure. Figures \ref{facc}c-d demonstrate the correct 
determination of $f_C = 1.5$ Hz at $P = 18.2$ kbar, whose measurement at fixed 
$f_C$ (a ``$T$-scan'') reproduces the same result as obtained from our $f$-scan 
analysis \cite{Larrea20}.

\begin{figure*}
\includegraphics[width=6.5cm]{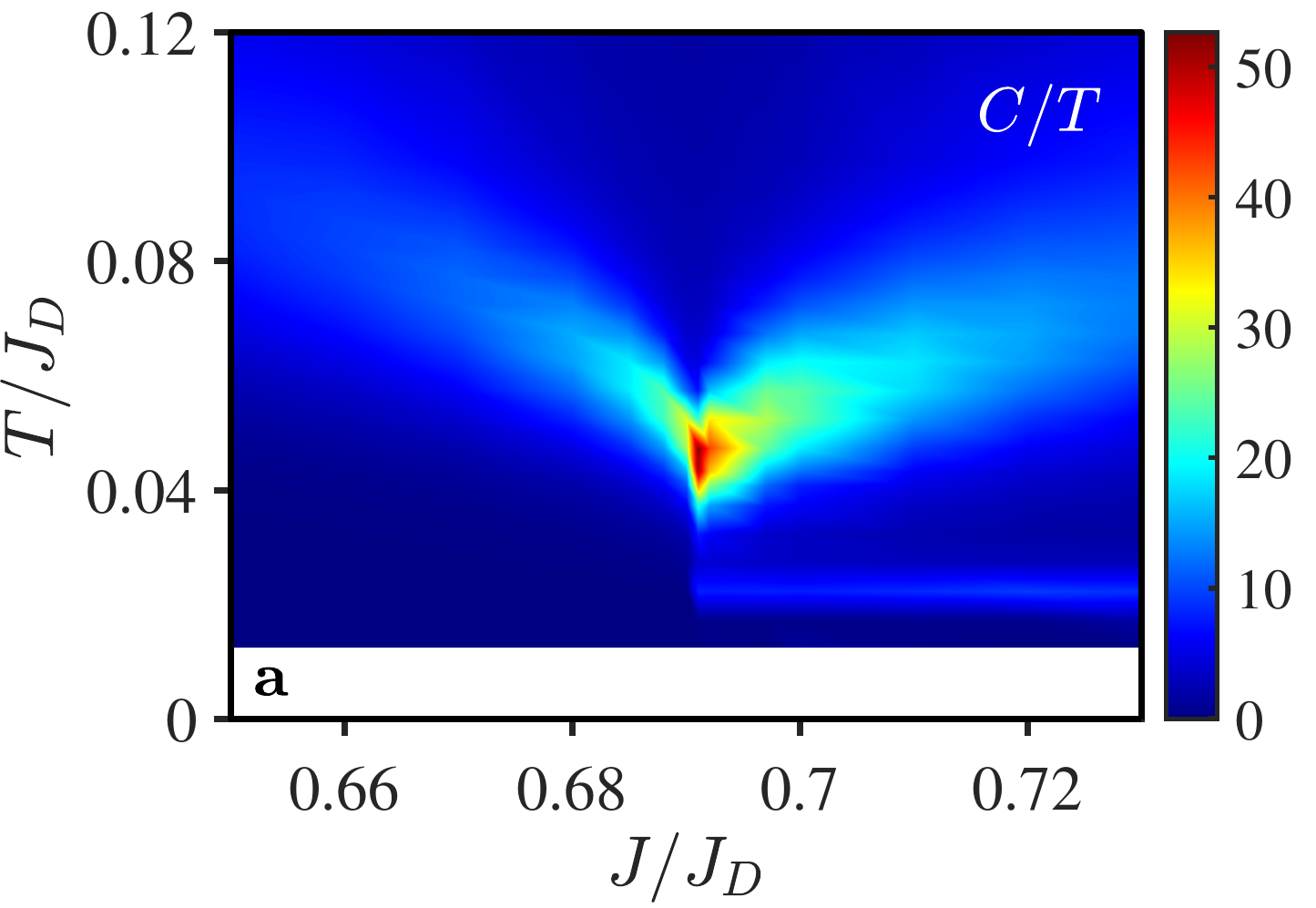}
\includegraphics[width=5.65cm]{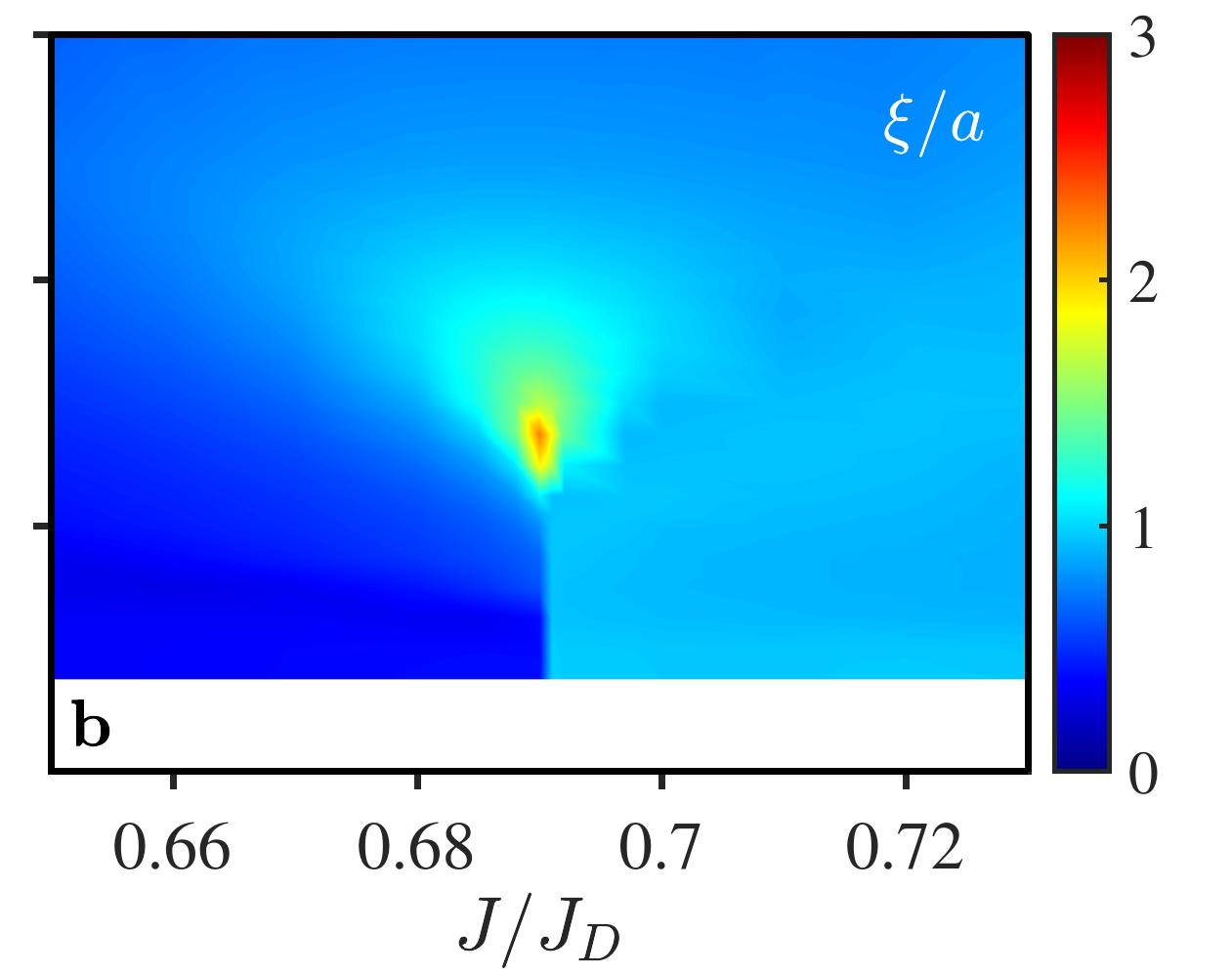}\hspace{-1mm}
\includegraphics[width=5.65cm]{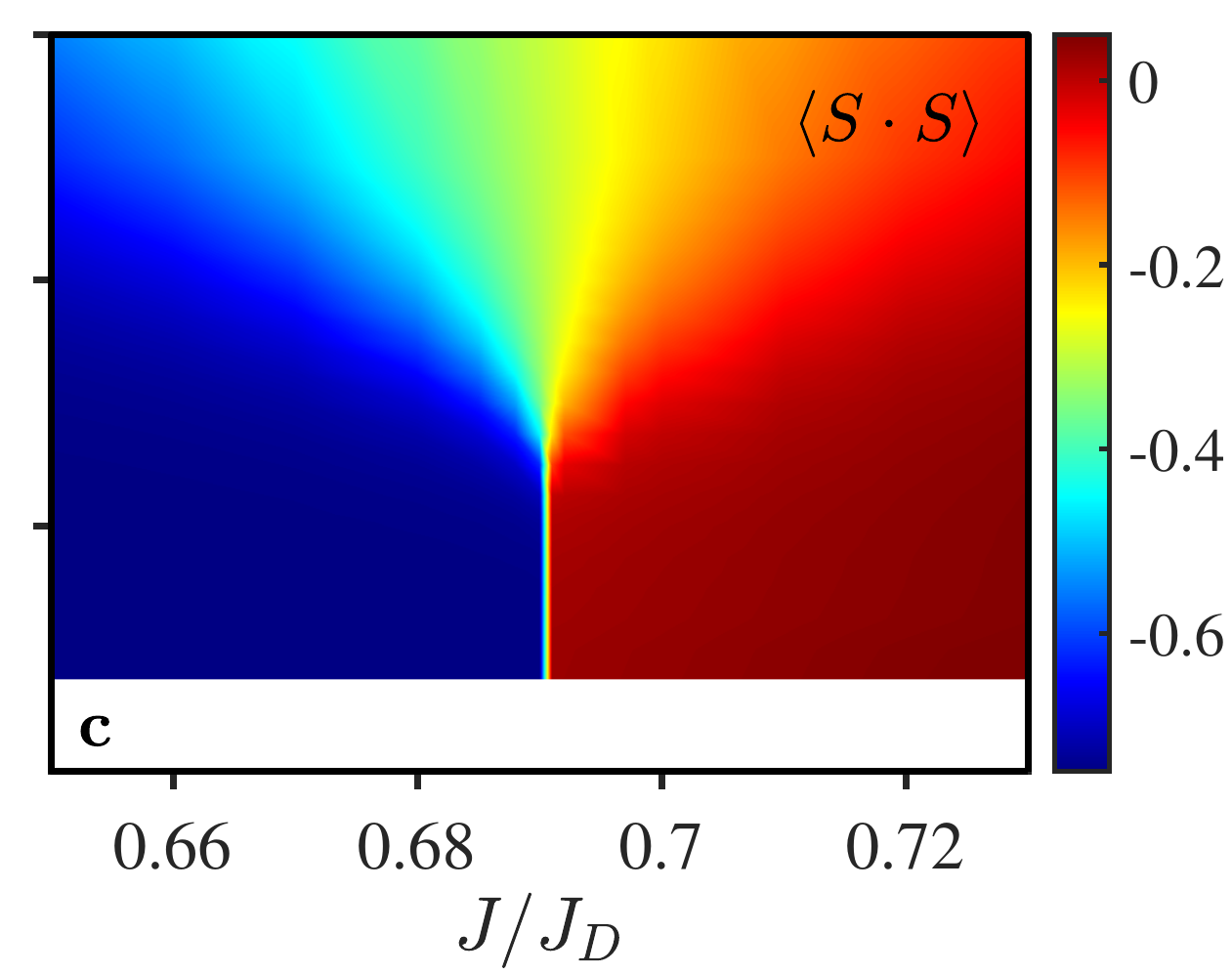}
\caption[]{{\bf The critical point in the presence of Dzyaloshinskii-Moriya 
interactions.} Thermodynamic data obtained from iPEPS calculations with 
$D = 10$ performed for the Shastry-Sutherland model in the presence of 
Dzyaloshinskii-Moriya interactions. These interactions are placed on the 
dimer ($J_D$) bonds and have the magnitude ($D_D/J_D = 0.03$) known for 
SrCu$_2$(BO$_3$)$_2$. They create an entangled ground state in the dimer 
phase, which resolves the numerical instabilities observed for the pure 
Shastry-Sutherland model at low temperatures, although the reduced symmetry 
limits the maximum $D$ to 10. {\bf a}, Specific heat, $C_p(J/J_D,T)/T$, shown 
in the same format as for Figs.~1b-c. {\bf b}, Correlation length, $\xi$, 
showing clearly the region of ``pressure'' and temperature over which Ising 
correlations develop. {\bf c}, Dimer spin-spin correlation function, $\langle 
{\vec S}_i \cdot {\vec S}_j \rangle$, emphasizing the abrupt onset with 
decreasing temperature of a sharp discontinuity as a function of $J/J_D$. 
It is clear that these Dzyaloshinskii-Moriya interactions have no qualitative 
effect whatsoever on the physics of the critical point.} 
\label{fig:CpDM}
\end{figure*}

For practical purposes in our measurements of SrCu$_2$(BO$_3$)$_2$, we found 
that the range of pressures and fields covered by our current investigation 
had negligible influence on the relevant parameters in the steady-state
equation. Thus we measured the $T$-dependence of the heat capacity of each 
sample at a constant field and at a fixed frequency, $f_C$, determined for 
each pressure. Our methodology allowed us to determine the heat capacity 
within an accuracy of 5\% with respect to an adiabatic technique. Further 
details concerning all aspects of the procedures of our a.c.~calorimetry 
measurements may be found in Ref.~\cite{Larrea20}. 

\medskip
\noindent
{\bf iPEPS.} Tensor-network methods provide a powerful approach for accurate 
numerical calculations of both the ground and thermal states of gapped local 
Hamiltonians. iPEPS \cite{Verstraete04,Nishio04,Jordan08}, a two-dimensional 
generalization of matrix-product states, are a variational Ansatz allowing  
both wavefunctions and thermal states \cite{Li11,Czarnik12,Czarnik15,
Kshetrimayum19,Czarnik19} to be represented efficiently in the thermodynamic 
limit, with the accuracy of the representation controlled systematically by 
the bond dimension, $D$, of the tensors. 

To obtain an iPEPS representation of a thermal state, we employ the algorithms 
of Ref.~\cite{Czarnik19}, which are based on the imaginary-time evolution of a 
purification of the thermal density operator. At each time step the bond 
dimension of the iPEPS is truncated to the maximal $D$. To maximize $D$ while 
working near the QPT, we restrict the truncation  to the (local) simple-update 
approach \cite{Jiang08,Wietek19}. Physical expectation values are computed by 
contracting the tensor network, which we perform by a development 
\cite{Corboz14} of the corner-transfer-matrix method \cite{Nishino96,Orus09}. 
This method is also used to compute the transfer matrix, from which we extract 
the correlation length, $\xi$ in Figs.~1d and \ref{fig:CpDM}b, of the 
correlation function with the slowest decay. To improve the efficiency 
of the calculations we exploit the global U(1) symmetry of the model 
\cite{Singh11,Bauer11}, which for a pure Heisenberg model is preserved in 
the presence of an applied field at finite temperatures.

\begin{figure*}
\includegraphics[width=5.9cm]{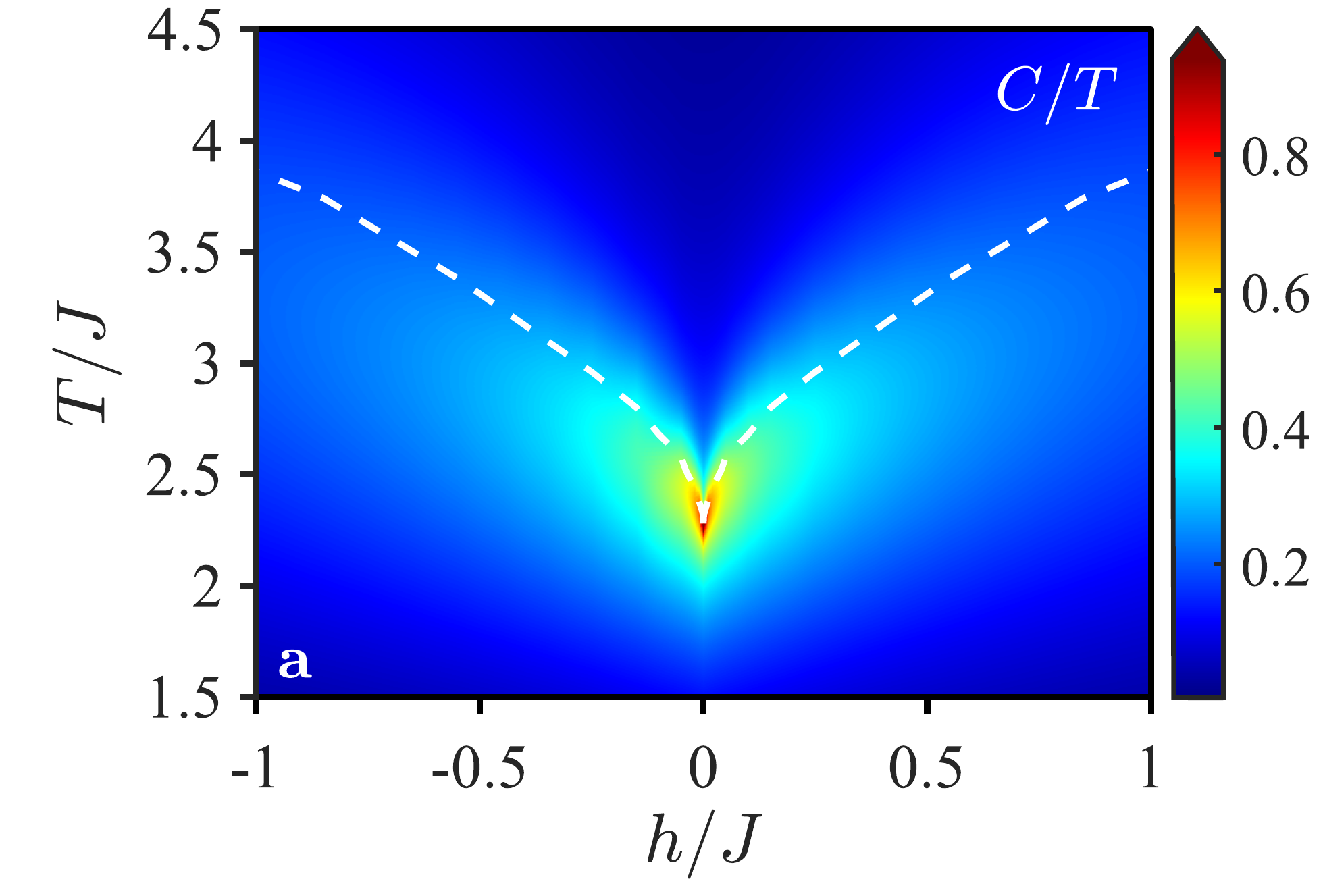}
\includegraphics[width=5.9cm]{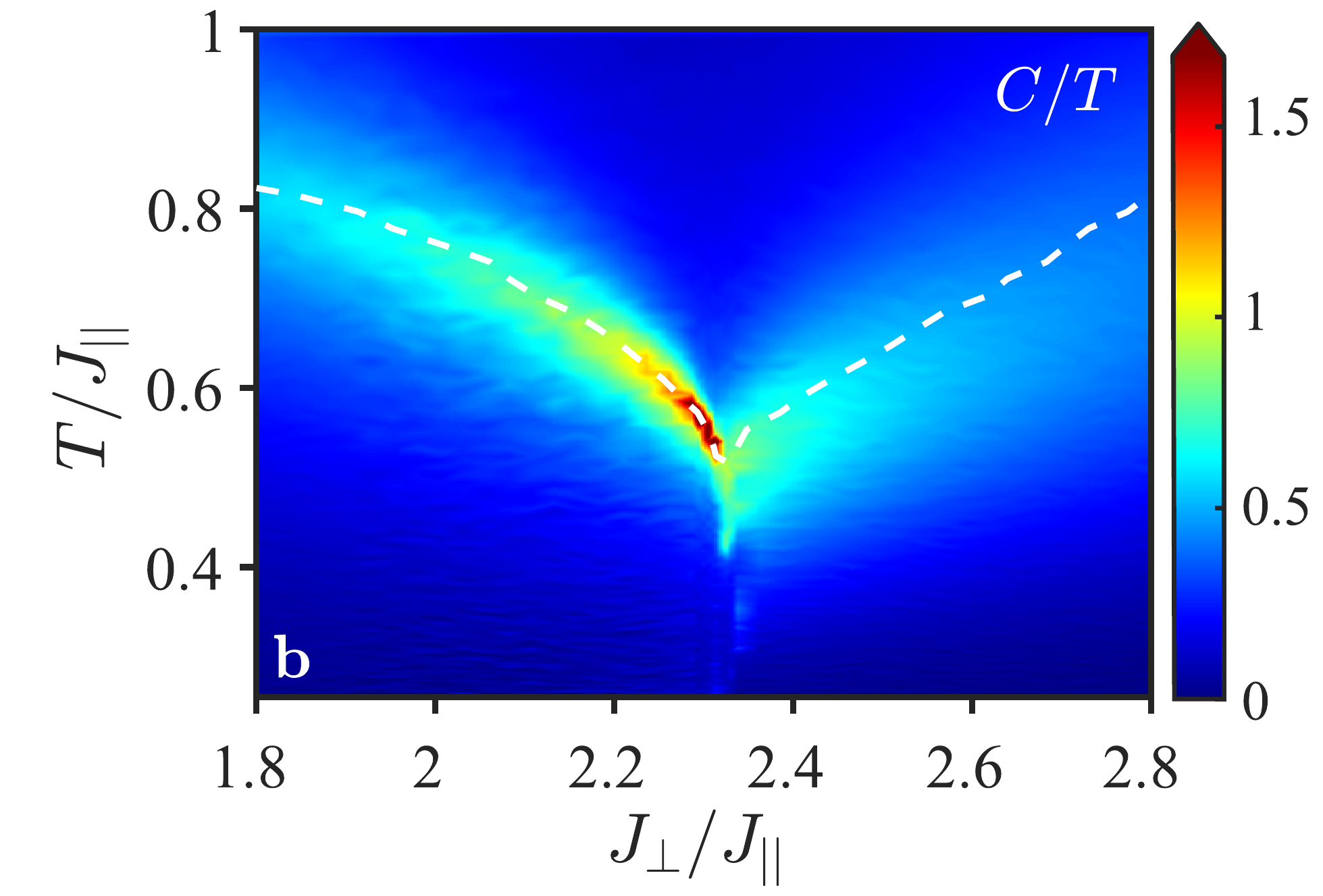}
\includegraphics[width=5.9cm]{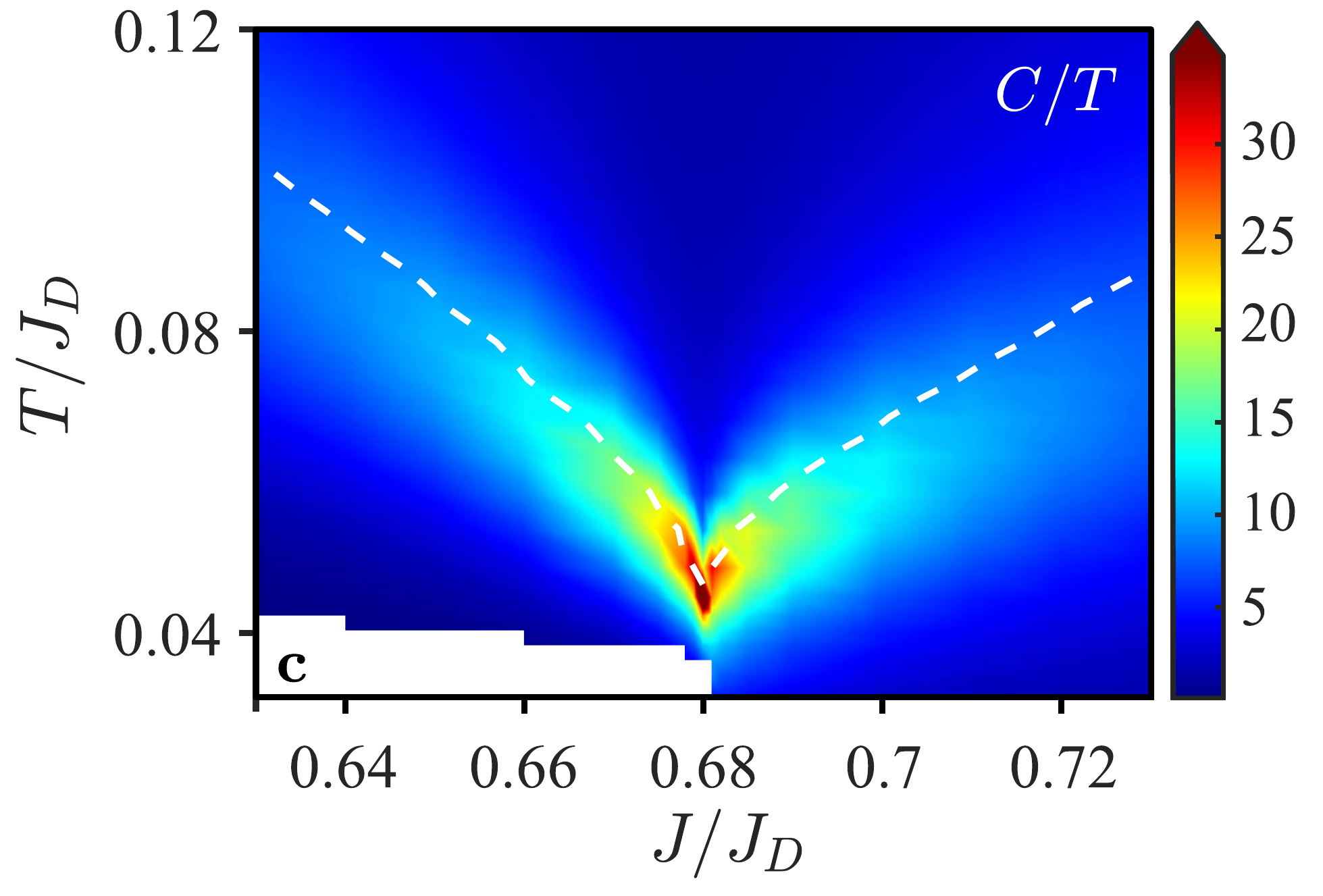}
\caption[]{{\bf Ising critical points in different lattice models.} 
Specific heat, $C/T$, for a number of 2D models, illustrating its universal 
behaviour around the Ising critical point. 
{\bf a}, Ising model on the square lattice in a longitudinal magnetic field, 
$h$, obtained by contracting the exact $D = 2$ tensor-network representation 
of the partition function using the CTM method with a boundary bond dimension 
$\chi = 24$ \cite{Nishino96}.
{\bf b}, Fully frustrated bilayer model, obtained by using the 
stochastic series expansion quantum Monte Carlo approach developed in 
Refs.~\cite{Stapmanns18,Wessel18} to perform simulations on systems of 
sizes up to 2$\times$32$\times$32 as a function of $J_\perp/J_\|$.
{\bf c}, Shastry-Sutherland model, obtained by iPEPS calculations with 
$D = 20$ as in Fig.~1c.
The dashed lines show the positions of the local maxima of the specific heat, 
$C (J/J_D)$, which we label by their temperatures, $T_{\rm max}$. These two 
lines reach an absolute minimum, $T_{\rm max} = T_c$, where they meet at the 
Ising critical point, with $T_{\rm max}$ increasing as the control parameter 
is changed away from the QPT. Thus the specific heat defines two characteristic 
lines in the phase diagram of the Ising critical point, rather than the single 
line given by the correlation length (Fig.~1d) and the critical isochore 
(Fig.~1e). We stress that this contrast is a fundamental property of the Ising 
model, and hence of all models sharing its physics. For the Shastry-Sutherland 
model ({\bf c}), the two lines of maxima could be taken to provide a 
qualitative definition of ``dimer-like'' and ``plaquette-like'' regimes, 
accompanied by a third regime bearing no clear hallmarks of either $T = 0$ 
phase. We remark that the values of $T_c$ in units of the relevant energy 
scale, $T_c/J = 2.3$ ({\bf a}), $T_c/J_\| = 0.53$ ({\bf b}) and $T_c/J_D = 0.04$ 
({\bf c}), vary widely among the three models. This can be traced to the change 
in slope of the ground-state energy at the transition, whose compensation by 
entropy effects restores a derivable free energy at $T_c$.}
\label{fig:CpminT}
\end{figure*}

In the Shastry-Sutherland problem, the optimal tensor geometry depends on the 
ground state \cite{Corboz13}: the spin correlations of the dimer phase are 
represented most efficiently using one tensor per dimer in a two-dimer unit 
cell \cite{Wietek19}, while correlations in the plaquette phase are best 
described by an Ansatz with one tensor per four sites (from four different 
dimers) on a plaquette \cite{Corboz13}. For working around a critical point, 
it is essential to use a single representation of both quantum phases, and 
in the Shastry-Sutherland model the plaquette basis is more efficient for 
representing the dimer phase than the reverse. For this reason, we have 
used the plaquette basis for all of the results reported here.

When working in the dimer phase, we do observe numerical instabilities at 
low temperatures, which are most probably an artifact arising due to the 
product nature of the dimer ground state, as all observables and energies 
are essentially converged at those temperatures. One means of circumventing 
this problem is to add a small Dzyaloshinskii-Moriya interaction to the 
model, in fact guided by the real spin Hamiltonian of SrCu$_2$(BO$_3$)$_2$. 
This leads to an entangled ground state, which as shown in Fig.~\ref{fig:CpDM} 
makes the low-temperature regime accessible, albeit at cost of a reduced $D$. 

When working in the plaquette phase, the strongly entangled nature of the 
plaquette ground state makes low-temperature studies very challenging, and 
for this reason our primary focus is on temperatures $T/J_D \gtrsim 0.03$, 
for a full characterization of the regime relevant to the critical point. 
Although a small feature is evident in $C/T$ around $T/J_D = 0.02$ 
(Fig.~\ref{fig:CpDM}a), we caution against associating this with the 
Ising transition of the plaquette phase, because we do not find corresponding 
behaviour in the spin correlations and thus cannot confirm the reliability of 
our results at such low temperatures.

\end{document}